\documentclass[aps,pre,twocolumn,superscriptaddress,showpacs,floatfix]{revtex4-1}
\usepackage{amssymb}
\usepackage{amsmath}
\usepackage{graphicx}
\usepackage{subfigure}	 	
\usepackage[english]{babel}
\usepackage{float}
\usepackage{subfigure}
\usepackage{color}
\usepackage{tikz}

\newcommand{\tikzcircle}[2][red,fill=red]{\tikz[baseline=-0.5ex]\draw[#1,radius=#2] (0,0) circle ;}
\usepackage{enumitem}

\usepackage{hyperref}
\hypersetup{
bookmarks=false,
pdftitle={RDM PS-2D},
anchorcolor=blue,
colorlinks=true,
citecolor=blue,
urlcolor=blue,
linkcolor=blue}
\usepackage[author=alguien]{pdfcomment}
\usepackage{xcolor}

\begin{document}

\title[Random dimer model in pseudo 2D lattices]{Random dimer model in pseudo
two-dimensional lattices}

\author{Uta Naether}
\email{naether@unizar.es}
\affiliation{Instituto de Ciencia de Materiales de Arag\'on and Departamento de
F\'{i}sica de la Materia Condensada, CSIC-Universidad de Zaragoza, 50009
Zaragoza, Spain}

\author{Cristian Mej\'{i}a-Cort\'es}
\affiliation{Departamento de F\'isica and MSI-Nucleus on Advanced
Optics, Center for Optics and Photonics (CEFOP), Facultad de
Ciencias, Universidad de Chile, Santiago, Chile}

\author{Rodrigo A. Vicencio}
\affiliation{Departamento de F\'isica and MSI-Nucleus on Advanced
Optics, Center for Optics and Photonics (CEFOP), Facultad de
Ciencias, Universidad de Chile, Santiago, Chile}

\begin{abstract}
In this work, we study long-time wave transport in correlated and uncorrelated
disordered 2D arrays. When a separation of dimensions is applied to the model,
we find that the predicted 1D random dimer phenomenology also appears in
so-called pseudo-2D arrays.  Therefore, a threshold behavior is observed in
terms of the effective size for eigenmodes, as well as in long-time dynamics.
For this threshold behavior to be observed a minimum system size is required,
what is very important when considering a possible experimental realization. For the long-time evolution, we find that for short-range
correlated lattices a super-diffusive long-range transport is observed, while
for completely uncorrelated disorder in 2D transport becomes sub-diffusive within the localization length and random binary pseudo-2D arrays show localization.
\end{abstract}



\maketitle

\section{Introduction}

The concept of wave localization due to disorder, known as Anderson
localization (AL), has been around for quite some time~\cite{A58}, and several
reviews have been written on this topic recently ({\it cf}.
Refs~\cite{50years,rev13a,rev13b}). This phenomenon appears as a consequence of
the destructive interference of multiple scattered waves and has been observed
in such different physical contexts as electronics, photonics, Bose-Einstein
condensates (see
Refs.~\cite{Alexp1,Alexp2,Alexp3,Alexp4,Alexp5,Alexp6,Alexp7}), to name a few. 
Whenever the corresponding physical system can be modeled with a Tight-Binding Hamiltonian with time-invariant potential for
non-interacting particles, AL can be found. This is particularly the case for the propagation of light in evanescently coupled optical waveguide arrays. Here, over recent years, impressive progress in experimental development in one (1D) and two dimensions (2D) has been made~\cite{SzaEXP,ValleEXP}. Diverse
quantum, and condensed matter, phenomena have been reproduced in these -clean
and macroscopic- setups, by using electromagnetic waves~\cite{Szagraph1,Szagraph3,bloch}. 
Moreover, it has been possible to incorporate controlled disorder during the fabrication of these photonic structures,
and different studies concerning  AL have been carried out~\cite{exprdm,rev13b}.

In general, localization properties depend strongly on the dimensionality of the
system~\cite{Kramac93}.  In difference to the three-dimensional (3D) case, in
1D and 2D already the slightest amount of uncorrelated disorder leads to a
complete exponential localization for all eigenmodes without any mobility edge,
even though the localization length is much larger for 2D systems~\cite{1D2D}.
But when correlations are included in the system, the picture changes
dramatically~\cite{delo} and long-range transport may be still possible, even
in low dimensions (for a recent review see~\cite{IKM12}).  The paradigmatic
example in 1D is the \textit{random dimer model} (RDM)~\cite{phillips,SAV10},
where the  pairing of adjacent on-site energies (dimers), at random positions
in the lattice, leads to two-site correlations for an otherwise random binary
model. For finite lattices of length $N$, the RDM shows that, below a certain
threshold of the disorder strength, there are $\sim\sqrt{N}$ extended (thus
transparent) states, resulting in super-diffusive wave-packet evolution below
the threshold, and diffusive transport exactly at the threshold region~\cite{phillips}. These
delocalized eigenstates were shown in experiments~\cite{exp,exp2}, whereas a
direct observation of the transport properties was reported only
recently~\cite{exprdm}.

A two-dimensional rectangular optical waveguide array can be thought as a classical analog to study quantum transport of two interacting particles in an one-dimensional chain, in the context of a Bose-Hubbard model. The problem is mapped to a 2D lattice, where the interaction between particles is described by taking the propagation constants at the lattice diagonal ($n=m$) different to the rest of the lattice~\cite{keil,krimer,lahini}. Recently~\cite{flachepl12}, it was shown that the interaction between particles could promote a metallic two-particle state in a 1D quasiperiodic lattice, whereas the single-particle (classical) regime presented no transport~\cite{aubry,aubryEXP}. Therefore, the study of a 2D random dimer lattice presents an interesting possibility to observe similar features in a model presenting a well defined transport transition.

In the present work, we will use a special construction of disorder in two
dimensions, called pseudo 2D, to map our system described by a tight-binding
Hamiltonian for non-interacting particles onto 1D chains. This enables us to
access the super-diffusive transport properties in such arrays. We will first,
in section~\ref{model}, present the model of pseudo-two-dimensional lattices
using a separation ansatz; then, in section~\ref{modes}, we use the extension of
eigenmodes in smaller lattices to show the threshold behavior and its dependence on system size. In section~\ref{evol} we show the super-diffusive transport in the numerical
long-time evolution of correlated arrays and compare the localization volume of
random binary pseudo-2D arrays to 1D lattices. Finally, in section~\ref{conclu} we conclude the present work.

\section{2D random dimer model}\label{model}

\begin{figure}[t]
\centering
\includegraphics[width=1\columnwidth]{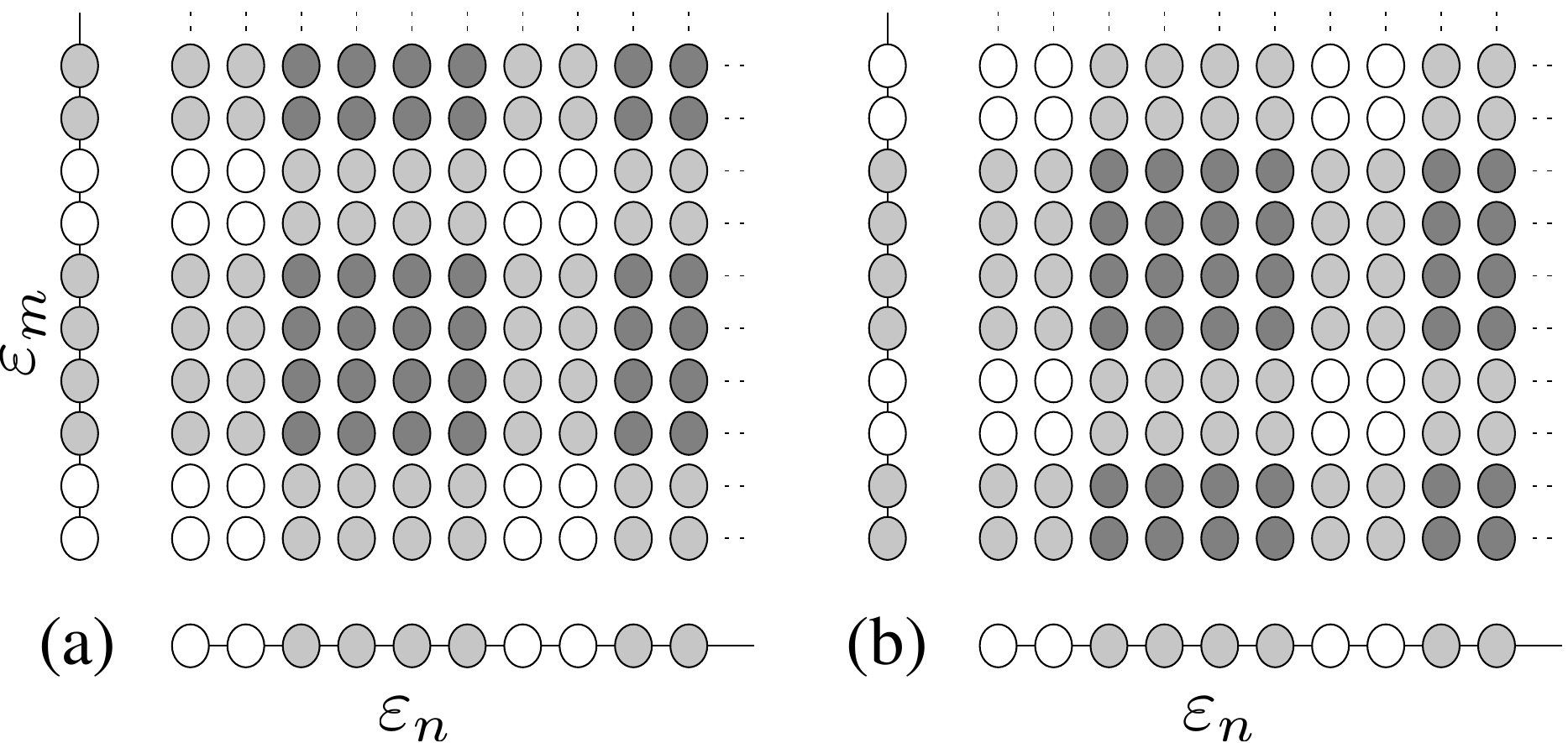}
\caption{Sketch of pseudo-2D random dimers: (a) {\sc ecoradi} and (b) {\sc dicoradi}. The individual realizations of $\epsilon_n$ and $\epsilon_m$ are shown on both axis. \tikzcircle[fill=white]{3.4pt} $\epsilon_{n,m}=0$, \tikzcircle[fill=lightgray]{3.4pt} $\epsilon_{n,m}=\Delta$, \tikzcircle[fill=gray]{3.4pt} $\epsilon_{n,m}=2\Delta$.}

\label{fig1}
\end{figure}

We start from a 2D discrete linear Schr\"odinger equation corresponding to a
tight-binding Hamiltonian for non-interacting particles, which also describes, e.g.,
the evolution of the field envelope amplitude of light propagating in the longitudinal direction $z$ in a 2D linear waveguide array~\cite{alex}:
\begin{equation}
\label{dls2d}
-i \operatorname{d}_z a_{n,m}=\epsilon_{n,m}a_{n,m}+\bigtriangleup_{nm}a_{n,m}\ , 
\end{equation}
with $\operatorname{d}_z\equiv \tfrac{\operatorname{d}}{\operatorname{d}z}$ and $\epsilon_{n,m}$ corresponding to the onsite-propagation constant at site $\{n,m\}$. Linear coupling, between different lattice sites of a square lattice, is defined as $\bigtriangleup_{nm}a_{n,m}\equiv C(a_{n,m+1}+a_{n,m-1}+a_{n+1,m}+a_{n+1,m})$, where $C$ represents the nearest-neighbor hopping constant. Without loss of generality, we set $C=1$, having in mind its rescaling effect on site energies and the effective propagation time/distance. Now, by restricting our study to the case $\epsilon_{n,m}\equiv\varepsilon_n+\varepsilon_m$, we make a dimension reduction by means of a separable ansatz: $a_{n,m}(z)=u_n(z)v_m(z)$. Thus, we obtain two \textit{independent} set of equations~\cite{alex} for each separable dimension:
\begin{eqnarray}
\label{eq1d2d}
\hspace{-1.5cm} -i \operatorname{d}_z u_{n}&=&\varepsilon_{n}u_{n}+ (u_{n-1}+u_{n+1})\ ,\\
 -i \operatorname{d}_z v_{m}&=&\varepsilon_{m}v_{m}+(v_{m-1}+v_{m+1})\ .\nonumber
\end{eqnarray}
The propagation constants $\varepsilon_l$ ($l=n,m$) are chosen in random pairs (dimers) as follows: 
\begin{equation}
\varepsilon_{l}=\varepsilon_{l+1}=\left\lbrace\begin{array}{c c c}
\Delta &,\ \mathrm{if} & \kappa \le 1/2\ ,\\
0 &,\ \mathrm{if} & \kappa >1/2\ ,
\end{array}\right.
\label{eq2}
\end{equation}
where $\kappa$ is chosen randomly in the interval $[0,1]$. $\Delta$ corresponds to the index (energy) contrast, which defines the differences in propagation constants (energies) between two different sites. In the following, we will consider four different cases of disorder realizations:
\begin{enumerate}[label=\roman{*}.]
\item Equally correlated random dimers ({\sc ecoradi}): $\varepsilon_n=\varepsilon_m$ [sketched in Fig.~\ref{fig1}(a)].
\item Different correlated random dimers ({\sc dicoradi}): $\varepsilon_n\neq\varepsilon_m$ [sketched in Fig.~\ref{fig1}(b)].
\item A binary case of uncorrelated random monomers ({\sc uncoram}); i.e., the case where the onsite propagation constants in model (\ref{dls2d}) are chosen randomly between two precise values: $0$ or $2\Delta$. 
\item A binary case of pseudo-2D uncorrelated random monomers ({\sc ramps}), where the onsite propagation constants {\it for each site} in model (\ref{eq1d2d}) are chosen randomly between the precise values: $0$ or $\Delta$.
\end{enumerate}
%
%

\section{Eigenmodes size}\label{modes}

To understand the fundamental properties of different lattices, it is crucial to get some information about the eigenmodes of every system. By analyzing their spatial localization features, we could gain a good insight of possible transport properties on the particular lattice. In order to investigate the spatial extension of a wave-packet in a square lattice of $N\times N$ sites with fixed boundary conditions, we define the normalized participation ratio as
\begin{equation}\label{R}
R\equiv\frac{P^2}{N^2\sum_{n,m}|u_{n,m}|^4} \ ,
\end{equation}
with $P=\sum_{n,m}|u_{n,m}|^2$ being a conserved quantity of model (\ref{dls2d}). In an optical waveguide array context, $P$ corresponds to the optical power (in other contexts as BEC's, this quantity is usually named as Norm or Number of particles). The participation ratio $R$ is a very useful quantity~\cite{diso1,stt}, which helps us to identify the number of effectively excited sites of a given profile, it can be understood as well as an effective occupied area of the profile. For a highly localized wave packet, $R$ approaches $1/N^2$, and tends to $1$ for the case of a completely homogeneous array excitation. We use it here to calculate the extension of the $N^2$ eigenmodes of a 2D square array with given disorder distribution.

\begin{figure}[t]
 \includegraphics[width=.45\textwidth]{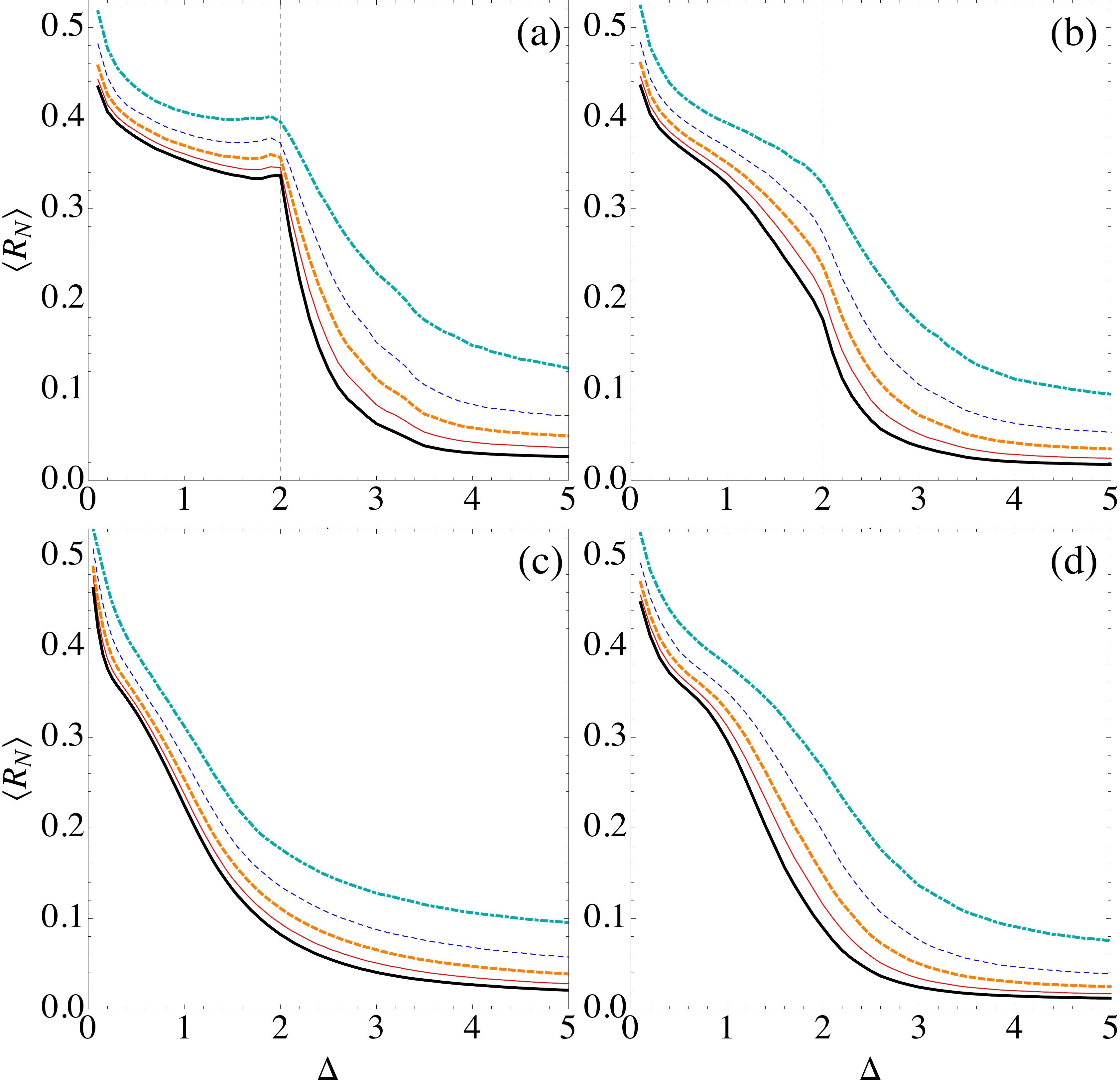}
\caption{$\langle R_N\rangle$ vs. $\Delta$ for (a) {\sc uncoram}, (b) {\sc dicoradi}, (c) {\sc uncoram} and (d) {\sc ramps} cases. All values are averaged over $100$ realizations of disorder. The  dot-dashed ( thin dashed, thick dashed, thin full, thick full) curve corresponds to $N=20, 30, 40, 50, 60$, respectively.}\label{fig2}
\end{figure}

Following the arguments presented in Ref.~\cite{phillips}, in an 1D array of $N$ lattice sites governed by the RDM, a fraction of $\sqrt{N}$ eigenmodes are extended over
the whole lattice, as long as $\Delta\leq2$. Thus, long-range transport is possible below a certain threshold. Therefore, in a separable pseudo-2D array described by Eqs.~(\ref{eq1d2d}), with system size $N\times N$, we expect the same behavior, but for $\sqrt{N^2}=N$ states. In
Fig.~\ref{fig2}, we plot $\langle R_N\rangle$ versus the contrast degree $\Delta$. $R_N$ is defined as the average participation ratio for the $N$ most extended (with largest $R$-value) eigenmodes, of a given realization and given $\Delta$. Then, $\langle R_N\rangle$ is obtained by averaging over $100$ realizations for each $\Delta$-value, for different system sizes ($N^2$). For this computation, we choose small systems sizes ($N:\{20,60\}$) to trace the appearance of the threshold behavior. This is done in order to estimate the smallest $N^2$--value required to observe the predicted phenomenology, which is important when thinking on a experimental implementation of the presented problem.
%

The participation ratio for all cases (i)-(iv) is displayed in Figs.~\ref{fig2} (a)-(d). For {\sc ecoradi} [see Fig.~\ref{fig2}(a)], the values of $\langle R_N\rangle$ first decrease to some kind of plateau until $\Delta=2$. In this region, the value of $\langle R_N\rangle>0.3$. This corresponds roughly to an average size of the eigenmodes larger than $30 \%$ of the lattice, what actually implies very delocalized states [for the ordered case ($\Delta=0$) and  for very small $\Delta$  extended linear modes  cover around $45-50 \% $ of the lattice]. For $\Delta>2$, $\langle R_N\rangle$ drops abruptly, especially for larger lattices. This phenomenon can be clearly attributed to the threshold behavior of the RDM~\cite{phillips,SAV10,exprdm}, what is highly dependent on the system size. In order to indeed observe more modes with a localization length larger than the extension of the lattice, a larger system size is required. From this figure, it is difficult to determine the emergence of the AL because this is a dynamical effect which is associated with the absence of diffusion across the lattice. However, we could identify that its appearance must occur for $\Delta>3$, where larger lattices have a $\langle R_N\rangle<0.1$. A rather similar scenario, but without a clear plateau and without such a pronounced threshold, is observed for the case {\sc dicoradi} [see Fig.~\ref{fig2}(b)]. For all system sizes, we observe a smooth decrement of $\langle R_N\rangle$, as a function of $\Delta$, with some change on the curvature around $\Delta\approx 2$ (a more abrupt decrement is observed after this region). For the {\sc uncoram} and  {\sc ramps} cases, the curves of $\langle R_N\rangle$ show a rather smooth and fast decrement, without any threshold signature; they show only a reduction of the mode participation ratio as disorder grows, as uncorrelated disordered systems must show. Furthermore, for the {\sc ramps} case we observe a smaller participation ratio for growing $\Delta$ than for the {\sc uncoram} case, suggesting a difference in localization volume for these cases, as will be confirmed in the following section \ref{evol}. For all the cases, we observe that for $N\gtrsim 50$ results converge, and that the transition threshold  becomes evident.

\section{Long-time Evolution}
\label{evol} 

To characterize and determine the diffusion of a wave-packet, we study the  evolution of the \textit{second moment}, which measures the size of a given profile in terms of its width with respect to a given center of mass. For two dimensional lattices, we compute this quantity for each dimension separately. First of all, we define the center of mass in the horizontal and vertical directions
%
\begin{eqnarray}
\label{smp}
\langle x(z)\rangle \equiv \frac{\sum_{n,m} n |u_{n,m}(z)|^2}{P}, \nonumber\\ 
\langle y(z)\rangle \equiv \frac{\sum_{n,m} m |u_{n,m}(z)|^2}{P},
\end{eqnarray}
%
respectively. Then, we define the second moments, for each dimension, as
%
\begin{eqnarray}
\label{sm}
M_x (z) \equiv \frac{\sum_{n,m} [n-\langle x(z)\rangle]^2 |u_{n,m}(z)|^2}{P}\ , \nonumber\\ 
M_y (z) \equiv \frac{\sum_{n,m} [m-\langle y(z)\rangle]^2 |u_{n,m}(z)|^2}{P}.
\end{eqnarray}
The evolution along $z$ is then averaged over the number of realizations ($L$) and over the horizontal and vertical directions. With this, we finally obtain an effective mean square displacement given by
\begin{equation}
\label{m2av}
\langle M(z)\rangle = \frac{1}{2L}\sum_{i=1}^L [M_{x}(z)+M_y(z)]_i\ .
\end{equation}

\begin{figure}[t]
\centering
 \includegraphics[width=0.46\textwidth]{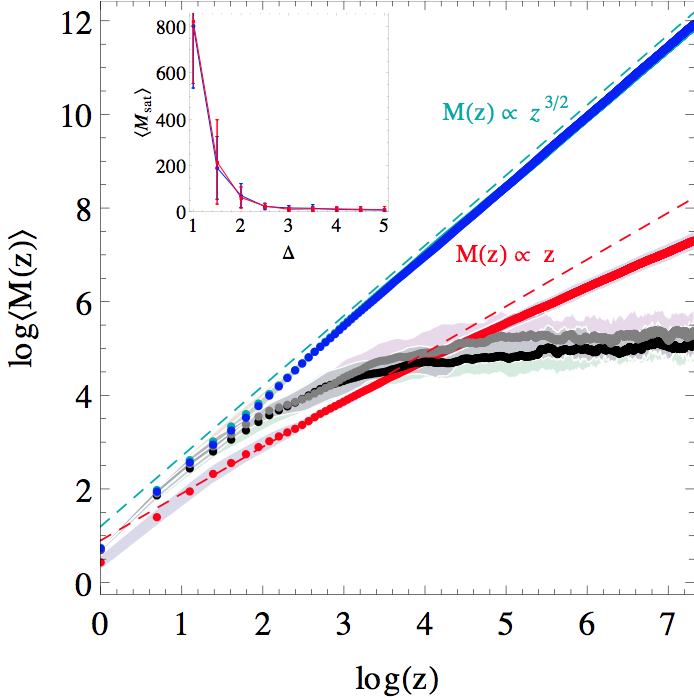}
\caption{$\log \langle M(z)\rangle$ vs. $\log z$, for  {\sc ecoradi} (green), {\sc dicoradi} (blue) and {\sc uncoram} (red). The dashed lines represent $M(z)=z^{\frac{3}{2}}$ (green) and $M(z)=z$ (red). {\sc ramps} for $\varepsilon_n=\varepsilon_m$ and $\varepsilon_n\neq\varepsilon_m$ are shown in black and gray, respectively. All curves are a result of averaging over $L=10$ realizations of disorder with $\Delta=1.5$, the shadows show the corresponding standard deviation. Inset: Saturation in random binary 1D arrays (blue) vs. {\sc ramps} lattices. (red)}\label{fig3}
\end{figure}

For 1D lattices, the mobility threshold is located at $\Delta= 2$~\cite{phillips, exprdm}. Below this threshold, the second moment evolves super-diffusively as $\langle M(z)\rangle \propto z^{3/2}$, while at the threshold the transport becomes diffusive with $\langle M(z)\rangle\propto z$. Above the threshold, the expansion is diffusive up to the localization length and then the second moment saturates, and diffusion is stopped. To prove the validity of the separation ansatz, we numerically integrate the original model equations (\ref{dls2d}) with the initial excitation $u_{n,m}(z=0)=\delta_{n,n_0}\delta_{m,m_0} $, located at the central site $(n_0,m_0)$ for an value of $\Delta=1.5$ below the threshold. To compute the long-time evolution up to $z_{max}=1500$, we implement a symplectic solver with a $SBAB_2$ integration scheme~\cite{symp}. The lattice sizes are chosen in such a way that the wave-packet spreading never reaches the border, reaching a maximum lattice extension of $N\times N=5120^2$ for the {\sc ecoradi} and {\sc dicoradi} cases, and $N\times N=4096^2$ for the {\sc uncoram} case. We average over $L=10$ realizations for each case. This may appear as an insufficient number of realizations at first glance; but, since every second moment, $M_{x}(z)$ and $M_{y}(z)$, includes the summation over the other dimension, curves are very smooth with nearly undistinguishable standard deviation, which is actually plotted with shadows for all cases in fig \ref{fig3}. 

Our main result, the evolution of $\log \langle M(z)\rangle$ for $\Delta=1.5$, is shown in Fig.~\ref{fig3} for {\sc ecoradi}, {\sc dicoradi}, {\sc uncoram} and two {\sc ramps} cases, in green, blue, red, black and gray, respectively.  We can confirm, that both the overlapping {\sc ecoradi} and {\sc dicoradi} evolve superdiffusively parallel to the green dashed line $3\log[z]/2$ showing no sign of change in curvature for growing $z$. Therefore, they behave as their one-dimensional counterparts \cite{phillips}. The case of {\sc uncoram} evolves still sub-diffusively (see that results are below the red dashed line). This can be understood by thinking about the localization length $l$ properties. In a completely random $2D$ array, $l$ scales exponentially~\cite{Alexp2} with the mean free path $\xi$ as $l\propto \xi \exp(\kappa \xi)$, in contrast to a $1D$ array for which $l\propto \xi$. In our case, the $N\times N=4096^2$ arrays are just too small to observe saturation of transport for a $2D$ array. Therefore, we also show the results for two kind of {\sc ramps} arrays, with $\varepsilon_n=\varepsilon_m$ (black) and $\varepsilon_n\neq\varepsilon_m$ (gray), that show earlier saturation, as expected for uncorrelated binary disorder. 

Since we observe the huge difference between localization length in the {\sc uncoram} and {\sc ramps} cases, we furthermore explored the localization behavior of {\sc ramps} vs. {\it real} 1D random binary arrays~\cite{bin}. In order to give a qualitative measure of the localization length, we define a mean value $\langle M_{sat}\rangle$, which is computed by averaging over the interval $z\in \{1000,1500\}$. The results are shown in the inset of Fig.~\ref{fig3}. We observe a saturation tendency for the random binary 1D array and  the 2D {\sc ramps} lattice tending to almost the same value of  $\langle M_{sat}\rangle$ for each contrast of disorder $\Delta$. By doing this, we demonstrate that the behavior of the pseudo 2D array is essentially one-dimensional, as was  suggested by the construction of the disorder. 


\section{Conclusions}\label{conclu}

Along this work we showed that a restriction of the distribution of on-site disorder facilitates the separation of dimensions in an originally two-dimensional array without interaction. This so-called pseudo-2D arrays show the same threshold behavior as the 1D case for a value equal to the sum of two 1D threshold values. If interaction would be included (leading to an effective nonlinearity), we would expect the threshold to increase in certain regions due to the renormalization of eigen-energies~\cite{flachepl12}. We found out that in order to observe a threshold behavior for a pseudo-2D lattice, a minimum system size is required $\sim 50\times 50$, what is crucial when thinking on the experimental implementation of this problem and observation of our findings. We also showed that, in a long-time evolution for short-range correlated arrays ({\sc ecoradi} and {\sc dicoradi}) and for $\Delta=1.5$, a super-diffusive long-range transport is observed, while for the {\sc uncoram} case transport becomes sub-diffusive and is expected to saturate for higher evolution times. The localization length behavior in the pseudo-2D random binary arrays is shown to be comparable to an one-dimensional lattice, which could be very useful to tune the localization volume in such setups. 
\section*{Acknowledgements}
The authors wish to thank the Spanish government project FIS 2011-25167,
FONDECYT Grants 1110142 and 3140608, Programa ICM P10-030-F, Programa de
Financiamiento Basal de CONICYT (FB0824/2008) and supercomputing infrastructure
of the NLHPC (ECM-02).


\end{document}